\documentstyle[12pt,aasms4,psfig]{article}

\lefthead{J.~Casares, P.A.~Charles, E.~Kuulkers}
\righthead{The mass of the neutron star in Cyg\,X-2 (V1341\,Cyg)}

\begin{document}

\title{The mass of the neutron star in Cyg\,X-2 (V1341\,Cyg)}

\author{Jorge Casares\altaffilmark{1},
	Philip Charles\altaffilmark{2},
	Erik Kuulkers\altaffilmark{2}}

\altaffiltext{1}{Instituto de Astrof\'\i{}sica de Canarias, 38200 La Laguna, 
Tenerife, Spain, jcv@ll.iac.es}

\altaffiltext{2}{Astrophysics, University of Oxford, Nuclear and
       Astrophysics Laboratory, Keble Road, Oxford OX1 3RH, United
       Kingdom, pac@astro.ox.ac.uk, erik@astro.ox.ac.uk}

\begin{abstract}

Cygnus X-2 is one of the brightest and longest known X-ray 
sources. We present high resolution optical
spectroscopy of Cyg\,X-2 obtained over 4 years which gives an improved
mass function of 0.69 $\pm$ 0.03\,M$_{\odot}$ (1$\sigma$).  In addition, we 
resolve
the rotationally broadened absorption features of the secondary star for
the first time, deriving a rotation speed of 
$v\sin i = 34.2 \pm 2.5$\,km\,s$^{-1}$ (1$\sigma$)
which leads to a mass ratio of 
$q = M_{\rm c}/M_{\rm x} = 0.34$ $\pm$ 0.04 (1$\sigma$, assuming a 
tidally-locked 
and Roche lobe-filling secondary).  Hence with the lack of 
X-ray eclipses (i.e. $i\lesssim 73^{\circ}$) we can set firm 
95\%\ confidence lower 
limits to the neutron star mass of 
$M_{\rm x} > 1.27$\,M$_{\odot}$
and to the companion star mass of 
$M_{\rm c} > 0.39$\,M$_{\odot}$. 
However, by additionally requiring that the companion must exceed
0.75\,M$_{\odot}$ (as required theoretically to produce a steady 
low-mass X-ray binary), then  
$M_{\rm x} > 1.88$\,M$_{\odot}$ and $i < 61^{\circ}$
(95\%\ confidence lower and upper limit, respectively), thereby making 
Cyg\,X-2 the highest mass neutron star measured to date. If confirmed this 
would set significant constraints on the equation of state of nuclear matter.

\end{abstract}


\section{Introduction}

The distribution of neutron star masses provide fundamental 
constraints on the equation of state of condensed matter. Very 
precise determinations (e.g., Thorsett et al.\ \cite{tho},
Nice, Sayer, \&\ Taylor \cite{nst96}) are available from 
time delays in millisecond radio-pulsars and are all consistent with 
1.38 $\pm$ 0.07\,M$_{\odot}$.
Dynamical masses can also be 
obtained for accreting neutron stars in X-ray binaries. In particular, 
high mass X-ray binaries (HMXBs) would appear to be ideal candidates because 
the orbit of the two components are 
measurable through spectroscopic doppler shifts and X-ray pulses. 
This, combined with a determination of the inclination angle through 
X-ray eclipses (when favourable) leads to a full solution for the 
system parameters. Following this strategy, six neutron star masses 
have been obtained, and they all lie in the range 1.0--1.9\,M$_{\odot}$ 
(see van Kerkwijk, van Paradijs \&\ Zuiderwijk \cite{kpz95} and references 
therein). However, the 
uncertainties involved in these determinations can be quite large due to 
non-Keplerian perturbations in the radial velocity curves. These are 
very difficult to assess and are caused by a variety of effects such as 
stellar wind contamination, tidal distortion of the companion 
and X-ray heating. 
On the other hand, mass determinations in low-mass 
X-ray binaries (LMXBs) are very difficult to obtain both 
because their neutron stars do not (usually) pulse and the
optical companions are normally overwhelmed by X-ray 
reprocessed radiation (van Paradijs \&\ McClintock \cite{vanmc}). Only in a 
few exceptional cases can
the companion be detected (when it is evolved or during X-ray off-states) 
and thus it becomes feasible to extract dynamical information 
and set constraints on the system parameters.

Cygnus\,X-2 is one of the few LMXBs in which the spectrum of the 
non-degenerate star is visible, contributing about 50 percent of the 
total visual flux. Estimates of the absolute magnitude of the donor 
and analysis of interstellar reddening (McClintock et al.\ \cite{mcc}) 
imply a distance of 
$\sim$8\,kpc and hence an X-ray luminosity of 
$L_{\rm x} \sim 10^{38}$\,erg\,s$^{-1}$. 
This large luminosity is consistent with near-Eddington accretion 
rates onto a neutron star as typically observed in LMXBs 
(e.g., Hasinger et al.\ \cite{has}); the neutron star's presence 
in Cyg\,X-2 is also indicated by the observation of X-ray bursts
(e.g., Kuulkers, van der Klis, \&\ van Paradijs \cite{kkp95}).
The X-ray intensity and energy 
distribution are highly 
variable on different time scales, tracing out a       
`Z' shaped track in the so-called X-ray colour-colour diagram with three 
distinct spectral states 
(Kuulkers, van der Klis, \&\ Vaughan \cite{kuu}, and references therein). 
Hence it is classified as a Z~source 
(Hasinger \&\ van der Klis \cite{hasvan}), the 
variations of which are believed to be triggered 
by mass transfer rate changes. Multi-wavelength observations 
indicate that the strength of the UV continuum and the high-excitation 
lines are correlated with the states of the `Z' diagram
(Vrtilek et al.\ \cite{vrt}; van Paradijs et al.\ \cite{van}).
Spectral type variations (in the range A5--F2) with orbital phase have been 
reported, with the earliest spectral type occurring when viewing the 
X-ray irradiated hemisphere of the companion
(Cowley, Crampton, \&\ Hutchings \cite{cow}).

From 1993 onwards we have collected high-resolution spectroscopy 
of Cyg\,X-2 with the aim of improving the system parameters and 
to resolve the rotation speed of the companion, thereby 
significantly refining the mass determination of the two components.
In addition, we have searched for the presence of Li in the 
atmosphere of the companion star (see e.g., Mart\'{\i}n et al.\ \cite{mart}), 
the results and implications of which will be presented elsewhere 
(Mart\'{\i}n et al., in preparation). 

\section{Observations}

We obtained 40 red spectra ($\lambda\lambda$6340--6800, 
0.40\,\AA ~pixel$^{-1}$ dispersion) 
of Cyg\,X-2 at the Observatorio del Roque de los Muchachos 
using the 4.2m William Herschel Telescope (WHT), equipped with the 
ISIS triple spectrograph (Clegg et al.\ \cite{cle}), 
on the nights of 1993 Dec 16--19,
1994 Oct 23--24, 1994 Dec 25, 1996 Aug 5, 1996 Dec 3, and 1997 Aug 1--7.
A 0.8--1.3 arcsec slit was used, depending on seeing conditions, 
giving spectral resolutions of 25--37\,km\,s$^{-1}$.
Cu-Ne arc spectra were obtained 
after every 1800\,s exposure of the target. For the sake of the spectral 
classification and rotational broadening analyses we also observed a 
grid of 34 template stars using exactly the same spectral configuration 
(with the narrowest 0.8 arcsec slit) as for Cyg\,X-2. These stars cover 
a range of spectral types from A0 to F8 in luminosity classes III, IV and V. 
   
\section{Results}

Individual radial velocities were extracted through cross-correlation of 
the red spectra with the template star HR\,2489 
(A9\,III), after masking out the broad H$\alpha$ and He\,{\sc I} $\lambda$6678 
emission lines. A subsequent sine wave fit to the velocity points 
(Fig.~1) gave the following parameters (after renormalising the minimum
reduced $\chi^2_\nu$ to 1): $P=9.8444 \pm 0.0003$\,d; 
$\gamma=-209.6 \pm 0.8$\,km\,s$^{-1}$; 
$K_{\rm c}=88.0 \pm 1.4$\,km\,s$^{-1}$; 
$T_0 ({\rm HJD})=2449339.50 \pm 0.03$, where 
$T_0$ corresponds to the standard zero phase definition, i.e., inferior 
conjunction of the secondary star. These and all subsequent errors 
quoted are $\pm$1-$\sigma$. In order to explore any non-symmetric effects in 
the radial velocity curve, e.g., artificial eccentricity induced by 
heating of the inner face of the companion (e.g., Davey \&\ Smith \cite{davey}),
we allowed the eccentricity $e$ to be a free parameter in the fit.
The presence of the eccentricity is only significant at the 
$\sim$75\%\ confidence
level; in this case we get $e=0.024 \pm 0.015$. Therefore, we 
conclude that a circular orbit represents the best description of the 
data points and thus we assume that our measured $K_{\rm c}$ corresponds 
to the true velocity semi-amplitude of the companion star. 
Our parameters are entirely consistent and substantially
more accurate than those derived by Cowley et al.\ (\cite{cow}) and 
Crampton \&\ Cowley (\cite{cra}).

Combining 
$K_{\rm c}$ and $P$ in the expression for the mass function gives 

\begin{equation}
{M_{\rm x} \sin^{3} i\over(1 + q)^2} = 0.69 \pm 0.03 {\rm M}_{\odot},
\end{equation}
 
\noindent
where $q=M_{\rm c}/M_{\rm x}$ is the system mass ratio.

Using the above ephemeris we obtained a doppler-corrected spectrum of
Cyg\,X-2, in the rest frame of the
secondary. A spectral type classification of A9 $\pm$ 2 for the companion
was then derived through two different techniques: optimal subtraction of
spectral type standards in the regions $\lambda\lambda$6380--6520,
$\lambda\lambda$6620--6665, $\lambda\lambda$6700--6760 (further details to
be found in Casares et al. \cite{cas}) and direct comparison of
the Fe\,{\sc I} line ratio $\lambda$6463/$\lambda$6457.
The former method was also applied to Doppler sums at the two
conjunction phases ($-$0.05 to 0.05 and 0.45 to 0.55),
but no spectral type variation could be found (Fig.2). Our result is
in contradiction with Cowley et al.\ (1979), who
claim orbital variations of the companion's spectral type
(due to X-ray heating) in the
range A5--F2. However, we note that their result is based on the ratio
Ca\,{\sc II}~K to H, which is not a good diagnostic because the Balmer series
(and perhaps also Ca\,{\sc II}~K) are clearly filled in by variable emission 
cores.
Indeed, it has been noted (Kristian, Sandage, \&\ Westphal \cite{kris};
Cowley et al.\ \cite{cow}) that the Balmer
lines in Cyg\,X-2 appear {\it abnormally broad}, which is expected if the
absorption cores are filled in with emission.
Therefore, we give more weight to the metal line ratios, which do not
support spectral variations larger than two subtypes between the two
conjunctions. We note that we actually detect an enhancement of the He\,{\sc I} 
$\lambda$6678 absorption 
at phase 0.5, together with an overall weakening of the 
metallic absorptions, produced by a $\sim$60 percent increase in the 
continuum. These are indications of heating effects,   
although spectral type variations are not significant since the relative 
depth of the metallic lines is maintained throughout the orbit. 
We also note that the absence of spectral type variations 
is not due to changes in the overall average X-ray luminosity
over the last $\sim$20\,yrs. In fact, recent 
{\it Rossi X-Ray Timing Explorer} (RXTE) All Sky Monitor  
measurements (Wijnands, Kuulkers \&\ Smale \cite{wks96}) indicate the overall 
X-ray luminosity to be 
comparable to that measured during the observations of Cowley et al.\ 
(1979).

In order to measure the rotational velocity of the companion star ($v \sin i$)
only the highest resolution spectra (with a 0.8 arcsec slit) were employed.
These correspond to the nights of 1996 Dec 3 and 1997 Aug 1--7. The
technique consists of performing a $\chi^2$ test on the residuals
after subtracting different broadened versions of our templates from
the doppler corrected sum of Cyg\,X-2. The template spectra were broadened
through convolution with the rotational profile of Gray (\cite{gray}) which
assumes a linearized limb darkening coefficient ($\epsilon=0.5$ at
6500\AA~ and $T_{\rm eff}=7000$\,K, see Wade \&\ Rucinski \cite{wadru}).
The broadened templates were also multiplied by a variable factor to
account for the continuum excess of the accretion disc, prior to
subtraction. Further details of this procedure can be found in, e.g.,
Marsh, Robinson, \&\ Wood (\cite{mar}).
Most of our templates are significantly broader than Cyg\,X-2 and
therefore only a subset of 20 templates (those with intrinsic
broadenings $\lesssim$15\,km\,s$^{-1}$) were considered in this
analysis. These all
give $v \sin i$ values for Cyg X-2 in the range 30--39\,km\,s$^{-1}$ with a 
mean of
$\sim$35.4\,km\,s$^{-1}$. The use of the most appropriate template
(A9~III) provides $v \sin i = 34.2 \pm 1.5$\,km\,s$^{-1}$ (see Fig.~3.)
The 1-$\sigma$ uncertainty in $v \sin i$ was derived by forcing
the minimum $\chi^2$ to increase by one
(Lampton, Margon \&\ Bowyer \cite{lam}), after renormalising the minimum
reduced $\chi^2_\nu$ to 1. 
However, we note that this error is purely formal and does not include 
sources of systematic error such as the uncertainty in the limb-darkening 
coefficient. More realistic estimates would be given by varying 
$\epsilon$ between its extreme values of 0 and 1. By doing this we find 
a more realistic $v \sin i = 34.2 \pm 2.5$\,km\,s$^{-1}$ which 
will be the value adopted by us hereafter.  
Because the donor star fills its Roche lobe and is synchronized with 
the binary motion, the rotational broadening provides a direct measurement 
of the binary mass ratio {\it q} through the expression
(e.g., Horne, Wade \&\ Szkody \cite{hor}):

\begin{equation}
v \sin i = K_{\rm c}~(1 + q)~\frac{0.49q^{2/3}}{0.6q^{2/3}+\ln{(1+q^{1/3})}}.
\end{equation}

Substituting our values of $v \sin i$ and $K_{\rm c}$ in Eq.~2 
we find $q=0.34 \pm 0.04$ which, combined with Eq.~1 gives 
$M_{\rm x} \sin^{3} i = 1.25 \pm 0.09$\,M$_{\odot}$.
On the other hand, the absence of X-ray eclipses provides a severe 
upper limit to the inclination of $i\lesssim 73^{\circ}$. 
This provides the following 95\%\ confidence
lower limits to the masses of the 
components in Cyg\,X-2: $M_{\rm x} > 1.27$\,M$_{\odot}$ and 
$M_{\rm c} > 0.39$\,M$_{\odot}$. 
The results are plotted in Fig.~4.

\section{Discussion}

Our conservative lower limit on $M_{\rm c}$ is well below the minimum 
secondary mass of $\sim$0.75\,M$_{\odot}$ required in recent 
theoretical predictions 
for steady LMXB sources (King et al.\ \cite{kin}). Imposing 
this condition of $M_{\rm c} \ge 0.75$\,M$_{\odot}$ now yields 
a 95\%\ confidence lower limit of $M_{\rm x} > 1.88$\,M$_{\odot}$ and 
a 95\%\ confidence upper limit of $i < 61^{\circ}$.
This would make Cyg\,X-2 the 
heaviest neutron star mass measured to date, and thereby provide support for 
stiff equations of state for nuclear matter 
(e.g., Cook, Shapiro, \&\ Teukolsky \cite{cook}) and would contradict the 
``softer'' equations of state as described in, e.g., Brown \&\ Bethe 
(\cite{bro}). On the other hand, if we assume a maximum possible
mass of the neutron star of $\sim$3.2\,M$_{\odot}$ (e.g., Rhoades \&\ Ruffini
\cite{rr74}),
we infer a 95\%\ confidence upper limit of $M_{\rm c} < 1.28$\,M$_{\odot}$ and 
a 95\%\ confidence lower limit of $i > 45^{\circ}$.
We note that the inclination constraints are more or less consistent
with recent ellipsoidal model fits to a compilation of BV photometric
light curves (Orosz \&\ Kuulkers, in preparation).

The RXTE has discovered a maximum 
kHz QPO frequency at 1066--1171 Hz in 8 persistent LMXBs which, if interpreted 
as the orbital frequency of the last marginally stable orbit, implies neutron
star masses of 2.0 $\pm$ 0.2\,M$_{\odot}$
(Zhang, Strohmayer, \&\ Swank \cite{zha}, see also Kaaret, Ford, \&\ Chen 
\cite{kfc97})). Recently 
kHz QPO have also been discovered in Cyg\,X-2 (Wijnands et al.\ \cite{whk97}),
so our mass estimate is in 
excellent agreement with that expected by Zhang et al.\ (\cite{zha}).
As they noted, this would be consistent with 
current evolutionary scenarios for LMXBs, where neutron stars would be 
born at 1.4\,M$_{\odot}$ but accrete at near-Eddington rates for 
$\sim$10$^8$ years (van den Heuvel \&\ Bitzaraki \cite{vanheu}). 
Dynamical mass determinations of other persistent 
LMXBs with evolved secondaries (e.g., GX\,1+4) will help to construct 
the distribution of neutron star masses and thereby allow 
new constraints to be set on the equation of state of nuclear-density matter.  

\acknowledgements

We thank Tom Marsh for the use of his optimal extraction 
routines and the MOLLY 
analysis package, and Tariq Shahbaz for discussions and assistance in 
preparing the figures. 
We also thank Vik Dhillon, Ren\'e Rutten and Miriam
Centuri\'on for supporting the Service observations, and Ian Browne, Neal 
Jackson and Peter Wilkinson for an exchange of observing time. JC acknowledges 
support by the EU grant ERBFMBI CT961756. 
The WHT is operated on the island of La Palma by the Royal Greenwich 
Observatory in the Spanish Observatorio del Roque de Los Muchachos of the 
Instituto de Astrof\'\i{}sica de Canarias.

\begin{figure}
\centerline{\hbox{
\psfig{figure=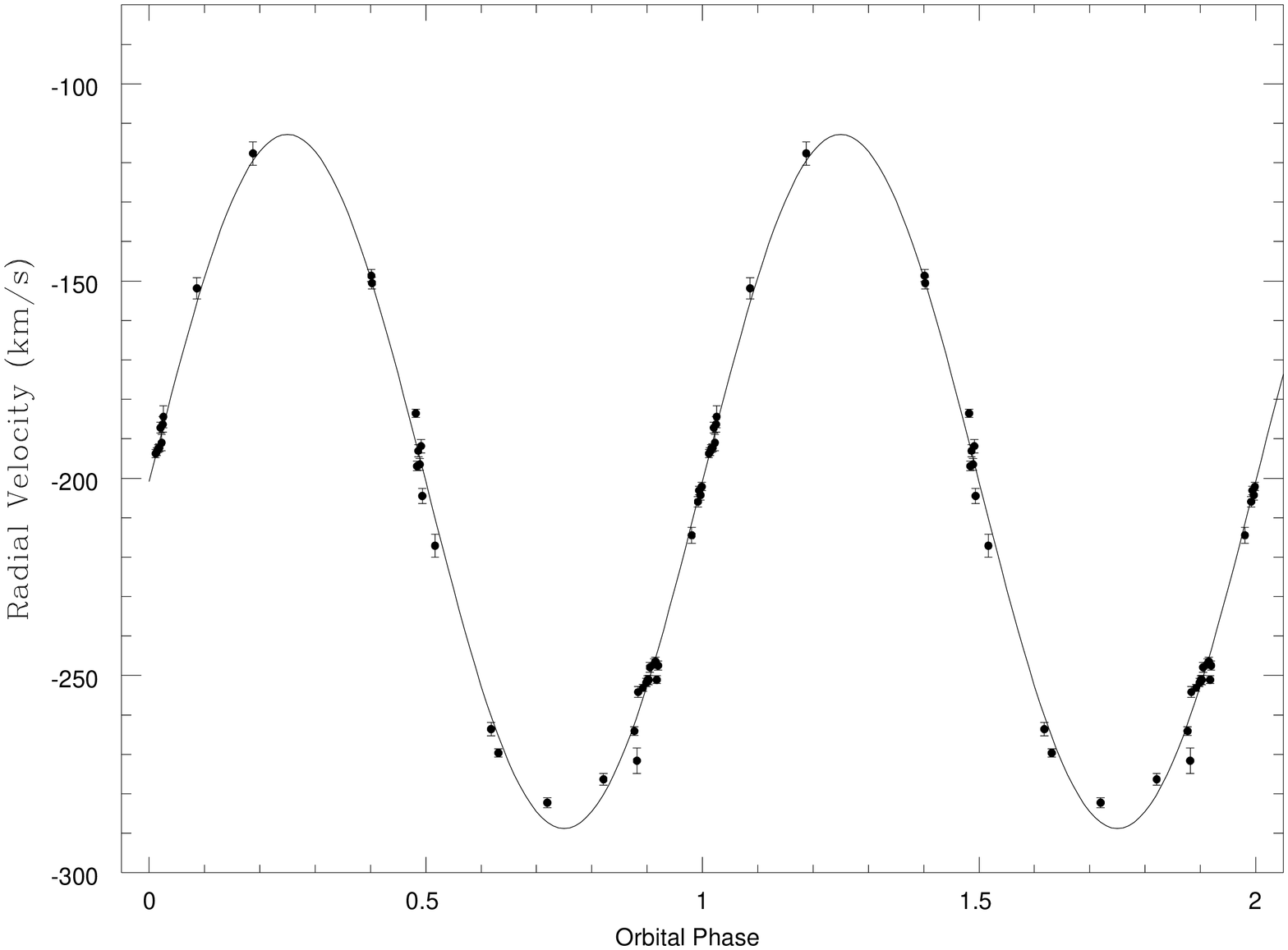,angle=-90,width=17cm}
}}
\caption{Radial velocity curve of the secondary star in Cyg\,X-2 folded 
on the ephemeris given in the text, plus the corresponding 
best sinusoidal fit.}
\end{figure}

\begin{figure}
\centerline{\hbox{
\psfig{figure=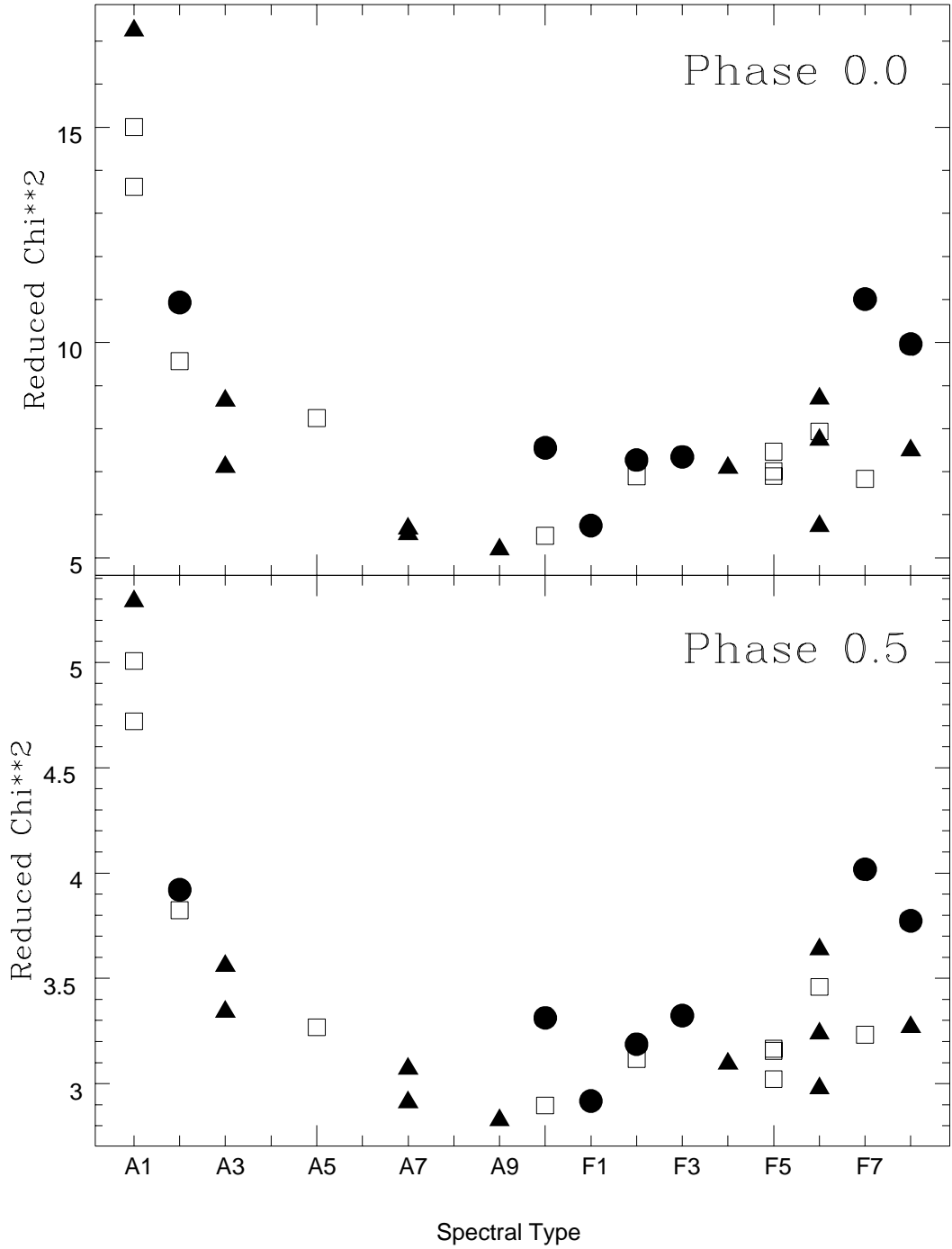,width=15cm}
}}
\caption{Reduced $\chi^2$ of the optimal subtraction technique 
as a function of the spectral type of the templates. 
Different symbols indicate different luminosity classes:
filled triangles for giants, open squares for subgiants and filled circles 
for main dwarfs.}
\end{figure}

\begin{figure}
\centerline{\hbox{
\psfig{figure=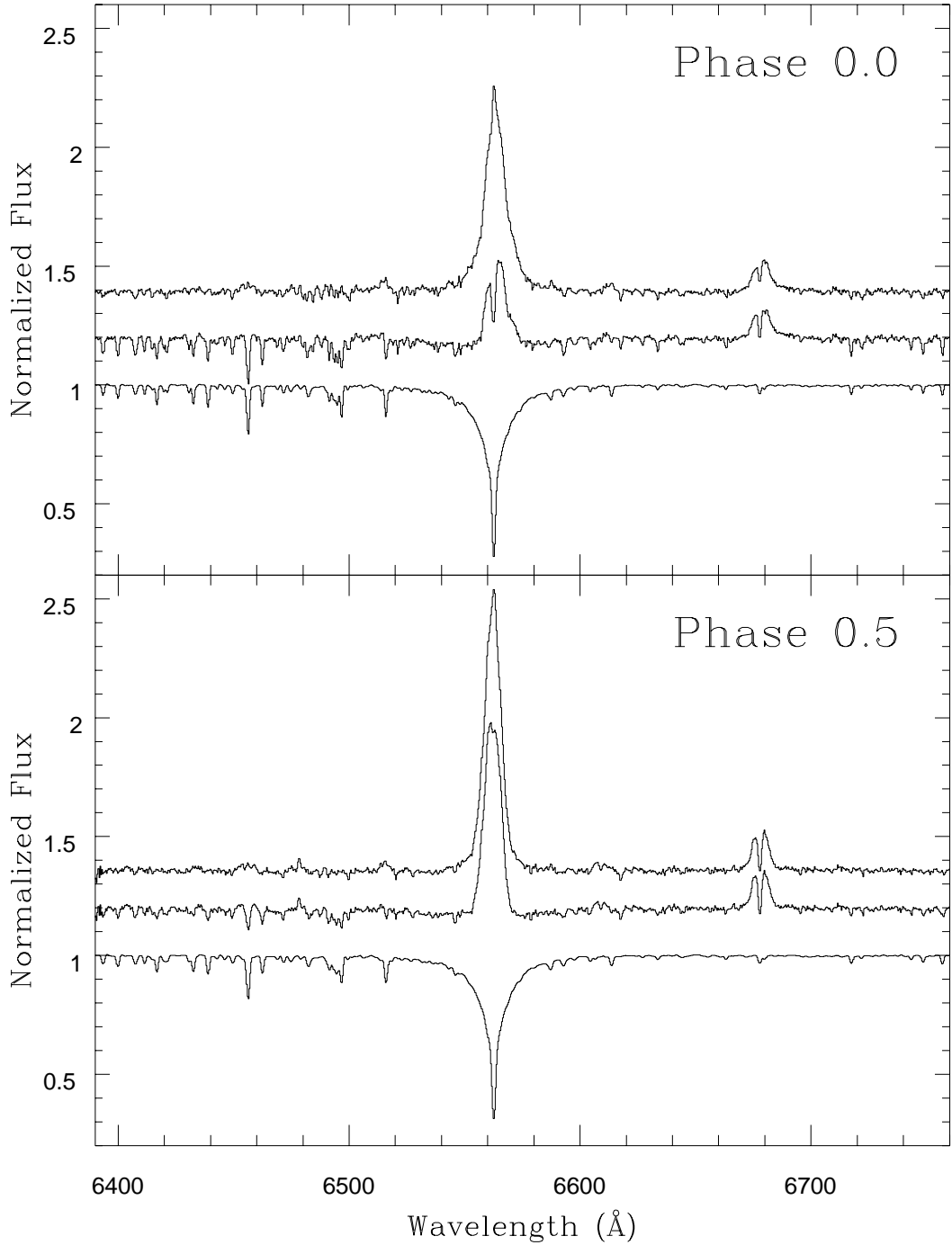,width=15cm}
}}
\caption{Normalised spectra of Cyg\,X-2 and the A9\,III template 
HR\,2489 at the 
two conjunction phases.
From top to bottom we plot the residual Cyg\,X-2 spectrum after subtraction 
of the broadened template, the 
doppler corrected sum of Cyg\,X-2 and the HR\,2489
spectrum broadened by 34\,km\,s$^{-1}$.}
\end{figure}

\begin{figure}
\centerline{\hbox{
\psfig{figure=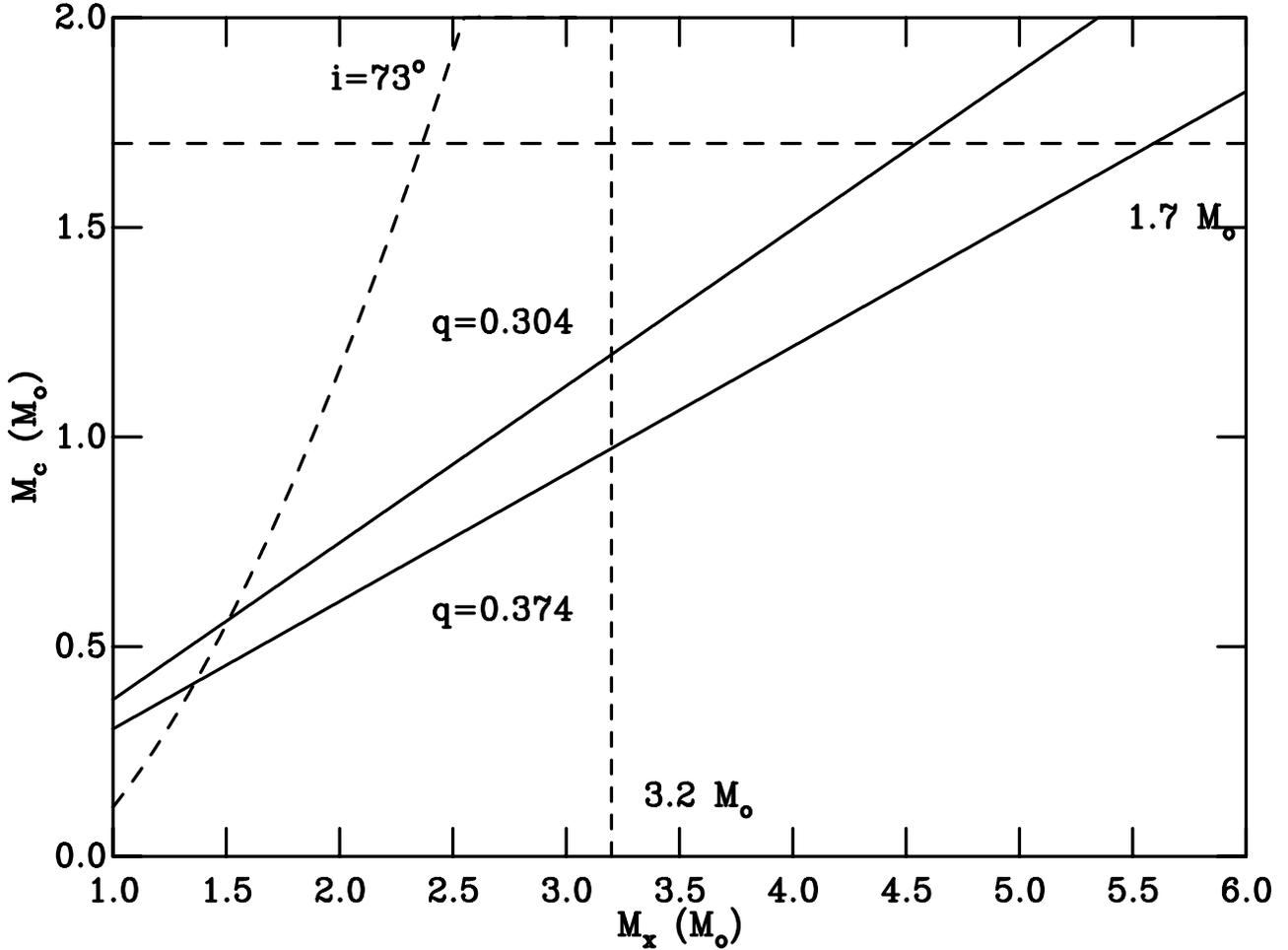,bbllx=79pt,bblly=83pt,bburx=526pt,bbury=673pt,angle=90,width=17cm}
}}
\caption{The mass of the companion ($M_c$) versus the mass of the compact
object ($M_x$). Values for $M_c$ and $M_x$ for Cyg\,X-2 lie between the two
solid lines which indicate the 1-$\sigma$ error in the mass ratio $q$. The
curved dashed line represents the constraint on the binary inclination by the
lack of X-ray eclipes ($i\lesssim 73^{\circ}$). The vertical dashed line
indicates the maximum possible mass for the neutron star (3.2\,M$_{\odot}$,
e.g., Rhoades \&\ Ruffini 1974). The horizontal dashed line indicates the mass of
a main sequence star (1.7\,M$_{\odot}$, Allen 1973) of the same spectral type 
as observed for the companion in Cyg\,X-2 (A9).}
\end{figure}

\end{document}